\documentclass[preprint,showpacs,preprintnumbers,amsmath,amssymb]{revtex4}

\usepackage{graphicx}
\usepackage{dcolumn}
\usepackage{bm}
\usepackage{ucs}
\usepackage[utf8x]{inputenc}
\usepackage{amsmath,amssymb}
\usepackage{bbm,xcolor}

\def\thetab{{\bar\theta}}
\def\etab{{\bar\eta}}
\def\alphad{{\dot\alpha}}
\def\betad{{\dot\beta}}
\def\alphah{{\hat\alpha}}
\def\betah{{\hat\beta}}
\def\gammah{{\hat\gamma}}
\def\deltah{{\hat\delta}}
\def\iim{\mathrm{i}}
\def\Dc{\mathcal{D}}
\def\Oh{\mathrm{O}}

\DeclareMathOperator{\Vect}{Vect}
\def\cc{\circledcirc}

\begin{document}

\preprint{SB/F/360-08}

\title{Noncommutative associative superproduct for general supersymplectic
forms}

\author{A. De Castro}
\author{I. Martín}
\author{L. Quevedo}
\affiliation{Departamento de Física, Universidad Simón Bolívar,\\ Ap.
89000, Caracas 1080-A, Venezuela}

\author{A. Restuccia}
\altaffiliation[Also at ]{Departamento de Física, Universidad Simón
Bolívar, Ap. 89000, Caracas 1080-A, Venezuela}
\affiliation{Max Planck Institut für Gravitationsphysik, Albert Einstein
Institut,\\ Am Mühlenberg 1, 14476 Golm, Germany}

\date{\today}

\begin{abstract} 
We define a noncommutative and nonanticommutative associative product for
general supersymplectic forms, allowing the explicit treatment of
non(anti)commutative field theories from general nonconstant string backgrounds
like a graviphoton field. We propose a generalization of deformation
quantization \emph{{à la Fedosov}} to superspace, which considers
noncommutativity in the tangent bundle instead of base space, by defining the
Weyl super product of elements of Weyl super algebra bundles. Super Poincaré
symmetry is not broken and chirality seems not to be compromised in our
formulation. We show that, for a particular case, the projection of the Weyl
super product to the base space gives rise the Moyal product for
non(anti)commutative theories.
\end{abstract}

\maketitle

\section{\label{Intro}{Introduction}}

During the last string theory revolutions we have learned that an undeniable
relationship between strings and noncommutative field theories exists. As
exemplified by the study of $D$--branes in presence of a finite constant
$B$--field \cite{Schomerus:1999ug, Douglas:1997fm}, it seems that geometry can
be noncommutatively deformed whenever stringy effects are relevant. In
particular cases like that of strings in a constant graviphoton background, the
deformation produced lies in the Graßmann sector of superspace whose undeformed
coordinates are characterised by anticommutators, giving raise to what has been
called non(anti)commutative field theories \cite{Seiberg:2003yz}.

Enormous effort has been made understanding both noncommutative and
non(anti)commutative field theories and their relation to string theory
\cite{Seiberg:1999vs,  de Boer:2003dn, Ferrara:2003xy, Klemm:2001yu,
Ivanov:2003te, Martin:2001zv, Ketov:2004qf, Ito:2006ig}. Most studies introduce
a Moyal product to deform the (super)field algebra of known theories, as this is
precisely the kind of geometry that results from string constant backgrounds.
The resulting theories usually break Lorentz symmetry and, in the
non(anti)commutative case, also either supersymmetry or chirality. The Moyal
product construction is very satisfactory when dealing with deformations of
Euclidean (super)spaces, but breaks down in curved geometry where it should be
replaced by a more general object like the Kontsevich product. As an example,
the deformed geometry of $D$--brane worldvolumes in curved backgrounds is given
to third order in the deformation parameter precisely by such product
\cite{Cornalba:2001sm}, which unfortunately can only be constructed order by
order in a series expansion. On the other hand, ordinary Moyal
product-noncommutative field theories suffer from bad causal behaviour when time
is treated as a noncommuting coordinate. 

More recently, in view of the limitations inherent to such methods, an
alternative noncommutative deformation of gauge theories based on the notion of
the Weyl bundle has been proposed \cite{Asakawa:2000bh, Martin:2001zv}. In these
works, authors use the powerful machinery  developed by Fedosov
\cite{Fedosov:1994zz, Fedosov:1996fu} which consists in a Weyl deformation of a
symplectic tangent space, where the symplectic structure plays the role of the
deformation parameter. The advantage of using this method is twofold: it allows
the description of physical situations where the background field is not
constant, and in addition spacetime global symmetries remain unbroken by the
deformation itself since it affects the tangent and not the base space. This
approach to deformation quantization has been used extensively to study string
theory D-branes in non-constant B-fields and more general curved backgrounds
\cite{Asakawa:2000bh}, supermembrane models \cite{Martin:2001zv}, the fuzzy
sphere \cite{Kishimoto:2001qe}, the canonical formulation of noncommuative
gauge theories \cite{FranciscoPena}, and BRST symmetry in quantum field
theories \cite{bordemann-2000-210}.

One may naturally ask if its possible to generalize Fedosov's ideas to describe
more general physical setups like non(anti)commutative field theories associated
to non-constant graviphoton backgrounds. At first sight one could consider an
atlas where the problem may be stated in terms of constant fields on each open
set, as the Darboux theorem is only valid in the neighbourhood of a point and
not globally, but we must then consider highly nontrivial matching conditions on
the overlapping of the sets.  It is therfore more convenient to directly develop
a non(anti)commutative product for general fields extending the known one for
constant fields. Also, from a quantum field theoretical point of view,
constructing a mechanism that deforms superspace but preserves super Poincaré
symmetry (or its Euclidean counterpart) is relevant in its own right. Our work
has precisely this goal. 

The deformation quantization of certain Poisson brackets in the context of
Batchelor supermanifolds using Fedosov's approach was first done by M.
Bordemann, \cite{Bordemann:1999ca, Bordemann:1996kd}. In the present paper we
are mostly interested in field theories involving fermionic fields, as we
explicitly discuss in section \ref{NonAntiSection}. Although the Batchelor
approach is an interesting one from a mathematical and physical point of view it
does not allow the introduction of fermionic fields, it is for this reason that
the Batchelor definition of supermanifolds is not followed in supersymmetric
field theory.

In this paper we extend Fedosov's deformation quantization to ${\cal N}=1$
superspace. We take a symplectic supermanifold and ordinary superspace as the
tangent and base space respectively, and propose a new associative Weyl product
which is \emph{non(anti)commutative}, meaning it is \emph{noncommutative} for
even objects that naturally fulfil commutation relations, and
\emph{nonanticommutative} for odd objects naturally satisfying anticommutation
relations. Such product endows the space of superfields with a superalgebra
structure. A key feature of our formulation is the preservation of super
Poincaré symmetry, a fact derived from a deformation operator acting exclusively
on tangent superspace variables. The super Poincaré algebra acting on base
superspace is therefore totally transparent to such deformation. We prove that
in certain cases, the projection of the Weyl (super)product on the base
supermanifold, results in the nonanticommutative product associated to nilpotent
Q-- and D--deformations such as those studied in \cite{Seiberg:1999vs,
Ferrara:2003xy, Klemm:2001yu, Ivanov:2003te}. And further explore more general
situations where the nonanticommuting parameter is actually a function of the
base supermanifold variables. This last subject is not even formulated in the
literature as a supersymmetric extension using Kontsevich methods.
\cite{Kontsevich:1997vb}.

\section{\label{wab}{Weyl Algebra Bundle}}
In this section we review Weyl algebra bundle theory basics
\cite{Fedosov:1996fu}. We start with a symplectic manifold $(M,\omega)$ of
dimension $2n$, where the two-form $\omega$ defines a symplectic structure  on
each tangent space $T_x(M)$, $x\in M$.

The \emph{formal Weyl algebra} $W_x$ is an associative algebra of formal series
over $\mathbbm{C}$ which stems from the symplectic space $T_x(M)$. Its elements
are defined by
\begin{equation}
a(y,\hbar)=\sum_{k,|\alpha|\ge0}\hbar^ka_{k,\alpha}y^\alpha,
\end{equation}
with $\hbar$ being the formal parameter, $y\in T_x(M)$ with coordinates
$(y^1,\ldots,y^{2n})$, and $\alpha=\alpha_1,\ldots,\alpha_{2n}$ a multi-index
such that $y^\alpha= (y^1)^{\alpha_1} \cdots (y^{2n})^{\alpha_{2n}}$. It is
possible to introduce an ordering prescription for the terms if we define the
\emph{degree} of the variables to be $\deg y^i=1, \deg \hbar=2$. The terms in
the
series can then be arranged by increasing degrees $2k+|\alpha|$.
A more friendly notation is
\begin{equation}
a(y,\hbar)=\sum_{k,p\ge0}\hbar^ka_{k,\mu_1\ldots\mu_p}y^{\mu_1}\ldots y^{\mu_p},
\end{equation}
where it is understood that $p=0$ terms correspond to the sum
$\sum_k\hbar^ka_k$.

The product of elements in this algebra is given by the \emph{Weyl product rule}
\begin{equation}
\begin{aligned}
a\circ b&=\left.\exp\left( -\frac{\iim \hbar}2\omega^{ij}
\frac\partial{\partial y^i}\frac\partial{\partial z^j}\right)
a(y,\hbar)b(z,\hbar)\right|_{z=y}\\
&=\sum_{k=0}^\infty \left( -\frac{\iim \hbar}2\right)^k
\frac1{k!}\omega^{i_1j_1}\ldots\omega^{i_kj_k}
\frac{\partial^ka}{\partial y^{i_1}\ldots \partial y^{i_k}}
\frac{\partial^kb}{\partial z^{j_1}\ldots \partial z^{j_k}},
\end{aligned}
\end{equation}
which is associative and independent of the particular choice of basis on the
tangent space $T_x(M)$. The algebras $W_x$ constitute the fibers of the
\emph{Weyl bundle} $W$ of \emph{formal Weyl algebras}, which is defined through
the union of all fibers
\begin{equation}
W=\bigcup_{x\in M} W_x.
\end{equation}
\emph{Differential} $q$-forms on $M$ with values in $W$ are sections of the
bundle $W\otimes \Omega^q$ defined by it formal series
\begin{equation}
a(x,y,\hbar)=\sum_{k,p\ge0}\hbar^k
a_{k,i_1\ldots 1_p, j_1\ldots j_q}(x)
y^{i_1}\ldots y^{i_p} dx^{j_1}\wedge\ldots
\wedge dx^{j_q},
\end{equation}
where the coefficients $a_{k,i_1\ldots 1_p, j_1\ldots j_q}(x)$ are covariant
tensors fields symmetric in $i_1\ldots i_p$ and antisymmetric in $j_1\ldots j_q$
depending on a point in $M$. They constitute an algebra
$C^\infty(W\otimes\Omega^*)$ under the usual Weyl product and the exterior
product of forms, with unit identically equal to 1. The total degree $\deg
a=2k+p$ corresponding to the variables $\hbar$ and $y^i$ establishes a
filtration within this algebra $W\otimes\Omega^* \supset W_1\otimes\Omega^*
\supset W_2\otimes\Omega^* \supset \cdots$. For $q=0$ we have
\emph{sections} simply as formal series
\begin{equation}
a(x,y,\hbar)=\sum_{k,|\alpha|\ge0}\hbar^k
a_{k,\alpha}(x)y^\alpha,
\end{equation}
with $a_{k,\alpha}(x)$ covariant symmetric tensor fields on $M$.  The bracket of
two forms $a\in W\otimes\Omega^{q_1}$
and $ b\in W\otimes\Omega^{q_2}$ takes the form
\begin{equation}\label{commutator}
[a\, ,\,b]=a\circ b-(-1)^{q_1q_2}b\circ a.
\end{equation}
A form $a$ will be called central if for any $b\in C^\infty(W\otimes\Omega^*)$
the commutator \eqref{commutator} vanishes. The \emph{center} of
$C^\infty(W\otimes\Omega^*)$ consists exclusively of forms not containing $y$,
which suggests a projection into the center by setting $y$ to zero. For
$a\in C^\infty(W)$ we will denote such projection by
\begin{equation}\label{proy}
\sigma(a)=a(x,0,\hbar).
\end{equation}
It is customary to define operators
that rise and lower the rank of differential forms: $\delta:\;
C^\infty(W_p\otimes\Omega^q) \rightarrow (W_{p-1}\otimes\Omega^{q+1})$ and
$\delta^*:\; C^\infty(W_p\otimes\Omega^q) \rightarrow
(W_{p+1}\otimes\Omega^{q-1})$, and respectively lower and rise the degree of
each term in the formal series. These fulfil the following properties
\begin{itemize}
\item $\delta^2=(\delta^*)^2=0$,
\item $\delta(a\circ\;b)=\delta(a)\circ b+(-1)^q\,a\circ \delta(b)$, for
$a\in\Omega^q$
\item $a=\delta\delta^{-1}a+\delta^{-1}\delta a+a_{00}$,
\end{itemize}
where $a_{pq}$ denotes the homogeneous part of Weyl degree $p$ and cohomological
degree $q$ of $a\in W\otimes\Omega^q$. Note that the last relation is similar to
the Hodge-Rahm decomposition, yet $\delta$ is a pure algebraic operator on $M$
as we can see from its realization in local coordinates
\begin{equation}
\delta a=dx^k\wedge\frac{\partial}{\partial
y^k}a=-\frac{\iim}{\hbar}[\omega_{ij}y^i dx^j,a].
\end{equation}
To define a connection on the Weyl bundle one recalls there exists a
symplectic connection on any symplectic
manifold that is torsion free and preserves the symplectic structure 
\begin{equation}
\partial_k \omega_{ij}-\omega_{nl}\Gamma^n_{mk}-\omega_{kn}\Gamma^n_{ml}=0.
\end{equation}
Using Darboux coordinates one can check that the \emph{connection symbols}
$\Gamma_{ijk}$ are completely symmetric $\Gamma_{[ijk]}=0$. The symplectic
connection can be used to define a connection on the bundle $W\otimes
\Omega^*$ by
its action on elements of $C^\infty(W\otimes\Omega^*)$
\begin{equation}
\nabla a=dx^i\wedge\nabla_i a\; ,
\end{equation}
in Darboux coordinates it reduces to 
\begin{equation}
\nabla a=da +\frac{\iim}{\hbar}[\Theta,a],
\end{equation}
where the connection symbols are $\Theta=\frac12 \Theta_{ijk}y^iy^jdx^k$.
The properties of the symplectic Weyl bundle connection are
\begin{itemize}
\item $\nabla(a\circ\;b) = (\nabla a)\circ b+(-1)^q\,a\circ(\nabla b)$, for
$a\in\Omega^q$,
\item for any $\phi\in\Omega^q, \nabla(\phi\wedge a)
=d\phi\wedge a+(-1)^q\phi\wedge \nabla a\;.$
\end{itemize}
More general covariant derivatives $\Dc$ on the bundle may be considered
by adding a one form $\gamma$ globally defined on $M$ with values in $W$,
\begin{equation}
\Dc a=\nabla a+\frac{\iim}{\hbar}[\gamma,a].
\end{equation}
With this at hand the \emph{Weyl curvature} may be defined as
\begin{equation}
\Upsilon=R+\nabla\gamma+\frac{\iim}{2\hbar}[\gamma,\gamma],
\end{equation}
where $R$ is the curvature associated to the connection $\nabla$. The Weyl
curvature
satisfies the Bianchi identity
\begin{equation}
\Dc\Upsilon=\nabla\Upsilon+\frac{\iim}{\hbar}[\gamma,\Upsilon]=0.
\end{equation}
Furthermore, for any section $a\in C^\infty(W\otimes\Omega^*)$
\begin{equation}
\Dc^2 a=\frac{\iim}{\hbar}[\Upsilon,a].
\end{equation}
In general, transitions on the bundle $T(M)$ will induce transitions on
$W\otimes\Omega^*$. The infinitesimal gauge transformations on elements of
$C^\infty(W\otimes\Omega^*)$ are expressed as automorphisms given by
\begin{equation}
a\rightarrow a+[a,\lambda],
\end{equation}
with infinitesimal $\lambda\in C^\infty(W\otimes\Omega^*)$. The corresponding
gauge transformations for the connections $\Dc$ are
\begin{equation}
\Dc\rightarrow\Dc+\Dc\lambda\; ,
\end{equation}
consequently,
\begin{equation}
\Dc a\rightarrow\Dc a+[\Dc a,\lambda].
\end{equation}

There exists a relation between the center and an Abelian subalgebra of the Weyl
bundle that maps the Weyl product to the Moyal product through the projection 
\eqref{proy}. To establish such relation one first defines \emph{Abelian
connections} $\Dc_A$ as connections whose Weyl curvature is a
central form in $C^\infty(W\otimes\Omega^*)$, i.e.,
\begin{equation}
[\Omega_A,a]=0\quad \forall a\in C^\infty(W\otimes\Omega^*). 
\end{equation}
An example of such connections may be expressed as
\begin{equation}
\Dc_A a=\nabla a +\frac{\iim}{\hbar}[\omega_{ij}y^i dx^j+r,a],
\end{equation}
with $\deg r\geqslant 3$. The \emph{Abelian subalgebra} is the set $W_A=\lbrace
a\in W /\Dc_A a=0 \rbrace$ associated to $\Dc_A$. There is a one-to-one
correspondence between the projections defined above \eqref{proy} and elements
of $W_A$. It is clear that for every $a\in W_A$ there is a projection projection
to the center
\begin{equation}
\sigma(a)=a_{00}.
\end{equation}
Now, given $a_{00}$ there is a \emph{unique} element $a\in W_A$ with such
projection. By means of this construction, from the Weyl product of two
elements $a,\; b$ of $W_A$ we obtain a \emph{globally defined} star product on
$M$ which coincides with the Moyal product when $\omega_{ij}$ is constant and
the symplectic connection is zero
\begin{equation}
\sigma(a\circ b)=a_{00}\star b_{00}.
\end{equation}

\section{Weyl Superalgebra Bundles}
The main purpose of this work is the generalization of the Weyl bundle
construction to superspace in order to study deformations of \emph{superfield}
algebras, which can then be used to extend supersymmetric field theories
expressed in the superfield formalism. Following closely the known construction
of Weyl bundle, we will need a \emph{symplectic structure} $\omega$ defined on a
\emph{supermanifold} $\mathcal{M}$. In the following section we give the first
steps toward such extension of deformation quantization, by defining this
concepts and introducing a new Weyl product of superfields.

\subsection{Symplectic supermanifolds}\label{ssm}
The structure underlying the concept of supermanifold \footnote{For an extensive
review on the subject of supermanifolds from a mathematical point of view see
\cite{Varadarajan:2004yz}. The generalization of supermanifolds aimed at more
physically inspired definitions can be found in \cite{Deligne:1999qp,
Deligne:1999ur}} is that of $\mathbbm{R}^{p|q}$ space, i.e. the topological
space $\mathbbm{R}^p$ endowed with a sheaf
$C^\infty[\theta^1,\theta^2,\ldots,\theta^q]$ of super $\mathbbm{R}$--algebras
freely generated over $C^\infty(\mathbbm{R}^p)$ by the Graßmann
($\mathbbm{Z}_2$--graded) algebra basis $\{\theta^1,\theta^2,\ldots,\theta^q\}$.
We will represent the set of \emph{superspace coordinates} by 
\begin{equation}
z^A=(x^m,\theta^\alphah),\qquad m=1,\ldots p,\quad \alphah=1,\ldots, q,
\label{RpqCoordinates}
\end{equation}
where capital indices consist of the total set of superspace indices, which
comprise lowercase $m,n,\ldots$ for standard spacetime directions, and
$\alphah,\betah,\ldots$ for doted and undoted Graßmann-odd indices. The
anticommuting (odd) variables $\theta^\alphah$ obey
\begin{equation}
\{\theta^\alphah,\theta^\betah\}=0,\quad
\alphah,\betah=1,\ldots, q.
\end{equation}

Like usual manifolds, that are locally isomorphic to $\mathbbm{R}^p$,
a \emph{supermanifold} $\mathcal{M}$ of dimension $p|q$ consists of a
topological space $M$ together with a sheaf of $\mathbbm{R}$--algebras called
the \emph{structure sheaf} $\mathcal{O}_\mathcal{M}$, such that $\mathcal{M}$
is locally isomorphic to $\mathbbm{R}^{p|q}$. 

The differential geometry involving supermanifolds is quite similar to the
classical one, having similar definitions and properties for super vector
bundles, connections and actions of super Lie groups. The most significant
difference arises from the existence of differential forms with arbitrary high
degree, which hinders a straightforward generalization of integration.
Nevertheless, after introducing the Berezinian, integration over supermanifolds
can be easily defined.

A \emph{super vector bundle} of \emph{rank} $p|q$ over $\mathcal{M}$ is a sheaf
of locally free $\mathcal{O}_\mathcal{M}$--supermodules $\mathcal{V}$ of
dimension $p|q$. It can be also defined as a fibre bundle over
$\mathcal{M}$ with typical fiber $\mathbbm{R}^{p|q}$, having as structure group
the super general linear group $GL(p|q)$. The first example of super vector
bundle is the tangent bundle of a supermanifold, which can be constructed from
the derivations of the superalgebra $\mathcal{O}_\mathcal{M}$ satisfying
\begin{equation}
D(fg)=(Df)g+(-1)^{p(D)p(f)}f(Dg),\qquad \forall f,g\in \mathcal{O}_\mathcal{M}.
\end{equation}
where $p$ defines the $\mathbbm{Z}_2$--grading of objects (0 for even, 1 for
odd). The $\mathcal{O}$--supermodule of derivations of
$\mathcal{O}_\mathcal{M}$ is a vector bundle of dimension $p|q$ called the
\emph{tangent bundle} $T(\mathcal{M})$, whose local basis consist of even
$\partial/\partial x^m$ and odd
$\partial/\partial\theta^{\alphah}$ derivations. Sections of the tangent bundle
are called \emph{vector fields} and are locally written as 
\begin{equation}
Y=Y^Ae_A=y^m\frac{\partial}{\partial x^m}
+\eta^{\alphah}\frac{\partial}{\partial\theta^{\alphah}}.
\end{equation}
The set of vector fields $\Vect(\mathcal{M})$ is a \emph{super Lie algebra}.

The \emph{cotangent bundle} $\Omega^1_\mathcal{M}$ of $\mathcal{M}$ is the
\emph{dual} of $T(\mathcal{M})$, whose basis we denote by
$dz^A=(dx^m,d\theta^{\hat{a}})$. One can interpret the duality pairing as an
inner product
$\langle\cdot,\cdot\rangle:T(\mathcal{M})
\otimes \Omega^1_\mathcal{M}
\longrightarrow\mathcal{O}_\mathcal{M}$, satisfying
\begin{equation}
\langle fY,g\omega\rangle=(-1)^{p(Y)p(g)}fg\langle Y,\omega\rangle,\qquad
\forall f,g\in \mathcal{O}_\mathcal{M}.
\end{equation}
In the local basis, such product can be obtained by linearity from
\begin{equation}
\left\langle\frac\partial{\partial z^A}, dz^B\right\rangle=\delta_A^B.
\end{equation}
The \emph{wedge product} of 1-forms is defined through
\begin{equation}\label{dzdz}
dz^A\wedge dz^B=-(-1)^{p(A)p(B)}dz^B\wedge dz^A    
\end{equation}
where $p(A)$ indicates the parity of an object with index $A$. One can linearly
extend this definition to the \emph{exterior product}
$\wedge:\Omega^k_\mathcal{M}\times
\Omega^l_\mathcal{M}\longrightarrow\Omega^{k+l}_\mathcal{M}$ of $k$-forms with
$l$-forms, providing the set of all forms with the structure of a superalgebra
which is called the \emph{exterior superalgebra}
$\Omega^*_\mathcal{M}\equiv\wedge^*\Omega^1_\mathcal{M}$. Note that
in contrast with the wedge product of forms, supercoordinates and superfield
components conform supercommutative superalgebras like
\begin{equation}\label{zz}  
z^Az^B=(-1)^{p(A)p(B)}z^Bz^A,
\end{equation}
as the concepts of cohomological degree of a form, and that of Graßmann parity
are independent.

The derivative
$d:\mathcal{O}_\mathcal{M} \longrightarrow \Omega^1_\mathcal{M}$ defined by
\begin{equation}
\langle Y, df\rangle=Y(f),
\end{equation} 
extends uniquely to the \emph{exterior derivative} $d=dz^A\partial/\partial
z^A$ on the exterior superalgebra and satisfies
\begin{equation}
d^2=0,\qquad d(\omega\chi)=d\omega\wedge\chi+(-1)^p\omega\wedge d\chi,
\qquad \omega\in\Omega^p_\mathcal{M}.
\end{equation}

From here onwards we will take $p=2n$ even and $q=4N$ representing the
physically relevant set of odd variables, thus studying only the supermanifold
$\mathbbm{R}^{2n|4N}$.  In $4|4$ dimensions, for example we have $A=(m,\alpha,
\alphad)$., i.e. $z^A=(x^m,\theta^\alpha,\thetab_\alphad)$, with $\alpha=1,2$,
$\alphad=\dot1,\dot2$. The notation we use is consistent with that of Wess and
Bagger \cite{Wess:1992cp} for differential forms in superspace.

A \emph{symplectic supermanifold} is a $2n|4N$ dimensional supermanifold
$\mathcal{M}$ endowed with a closed, non-degenerated two-form $\omega$ called
the \emph{symplectic superform}.
In coordinates, 
\begin{equation}
\omega=\omega_{AB}dz^A\wedge dz^B.  
\end{equation}
 
The Darboux theorem \cite{Rothstein:1990bj, Kostant:1975qe, Khudaverdian:2000zt}
states that there exist local coordinates on the supermanifold where
the symplectic superform $\omega$ has the following block shape
\begin{equation}
\omega_{AB}=
\begin{pmatrix}
\omega_{mn}&0\\
0&\omega_{\alphah\betah}
\end{pmatrix},
\label{DarbouxOmega}
\end{equation}
with Graßmann-even components having definite symmetry under index exchange
$\omega_{mn}=-\omega_{nm}$ and $\omega_{\alphah\betah} =
\omega_{\betah\alphah}$.
The
symplectic superform can be used to rise and lower indices.
Symplectic supermanifolds always admit a torsionless affine connection
preserving the symplectic superform, called \emph{(super)symplectic
connection}. We
will denote the connection by $\nabla$, and define its coefficients
$\Gamma^A_{BC}$ by its action on the tangent base
\begin{equation}
\nabla_{e_A}e_B=\nabla_A e_B=\Gamma^C_{AB}e_C.
\label{ConnectionCoefficients}
\end{equation}
Such coefficients can be found from the Leibnitz property of the covariant
derivative
\begin{equation}
\nabla_{Y_1}(\omega(Y_2,Y_3))=(\nabla_{Y_1} \omega)(Y_2,Y_3)
+\omega(\nabla_{Y_1}Y_2,Y_3)
+\omega(Y_2,\nabla_{Y_1}Y_3)
\end{equation}

The conditions for the vanishing of the torsion in superspace can be derived
from its very definition in terms of vector fields $Y_1$ and $Y_2$
\begin{equation}
T=\nabla_{Y_1}Y_2-\nabla_{Y_2}Y_1-[Y_1,Y_2],
\end{equation}
resulting in the following components
\begin{equation}
\begin{aligned}
&T^A_{\;\;mn}=\Gamma^A_{\;\;mn}-\Gamma^A_{\;\;nm} \\
&T^A_{\;\;m \alphah}=\Gamma^A_{\;\;m\alphah}-\Gamma^A_{\;\;\alphah m} \\
&T^A_{\;\;\alphah\betah}=\Gamma^A_{\;\;\alphah\betah}+\Gamma^A_{
\;\;\betah\alphah}
\end{aligned}\label{supertorsion}
\end{equation}

A torsion free affine connection associated to our symplectic superform is
defined by the equation 
\begin{equation}
\nabla_X \omega(Y_1,Y_2)=\nabla_X [\omega(Y_1,Y_2)]-\omega(\nabla_X
Y_1,Y_2)-\omega(Y_1,\nabla_X Y_2)=0
\end{equation}
In Darboux coordinates, the resulting conditions on the connection coefficients
\eqref{ConnectionCoefficients} are
\begin{equation}
\begin{aligned}
&\omega_{kl}\Gamma^k_{\;\;mn}+\omega_{nk}\Gamma^k_{\;\;ml}=0\\
&\omega_{\alphah\betah}\Gamma^\betah_{\;\;mn}
-\omega_{nk}\Gamma^k_{\;\;m \alphah}=0\\
&\omega_{\alphah\betah}\Gamma^\betah_{\;\;\deltah \gammah}
+\omega_{\gammah \betah}\Gamma^\betah_{\;\;\deltah \alphah}=0\\
&\omega_{mn}\Gamma^n_{\;\; \alphah\betah}
-\omega_{\betah \deltah}\Gamma^\deltah_{\;\; \alphah m}=0
\end{aligned}
\end{equation}

Following \eqref{supertorsion}, the torsionless components of the symplectic
connection are

\begin{align*}
&\Gamma_{lnm}=\Gamma_{nlm}
&& \text{a completely symmetric object}\\
&\Gamma_{l \alphah n}=\Gamma_{n \alphah l}
&&\text{a completely symmetric object}\\
&\Gamma_{l \alphah \betah}=\Gamma_{\betah \alphah l}
&& \text{antisymmetric in 1,2 positions but symmetric in 2,3}\\
&\Gamma_{\gammah\alphah\betah}=-\Gamma_{\betah\gammah\alphah}
&&\text{a completely antisymmetric object}
\end{align*}

\subsection{Weyl superalgebra bundle through a new associative superproduct}

We will consider a symplectic supermanifold $(\mathcal{M},\omega)$ locally
isomorphic to $\mathbbm{R}^{4|4}$ representing $N=1$ symplectic superspace, as
most results are easily generalized to extended superspace. Local coordinates of
$z\in\mathcal{M}$, and the components of a generic vector
$Y\in T_z(\mathcal{M})$ in tangent space, are to be denoted
$z^A=(x^m,\theta^\alpha,\thetab^\alphad,)$ and $Y^A=(y^m, \eta^\alpha,
\etab^\alphad)$ respectively, where $\theta^\alphah,\eta^\alphah$ are
Graßmann-odd. Without
loss of generality we take the symplectic tensor defining the symplectic
structure over fibers of $T(\mathcal{M})$ to have block diagonal
components $\omega_{AB}$ as in \eqref{DarbouxOmega}.

A \emph{Weyl superalgebra} $\mathcal{W}_z$ defined locally on the tangent
superspace $T_z(\mathcal{M})$ is an associative superalgebra over
$\mathbbm{C}$, its elements being formal series on the parameter $\hbar$
\begin{equation}
f(Y,\hbar)=\sum_{k,p}\hbar^k\; f_{k,A_1,\ldots,A_p} Y^{A_1}\cdots Y^{A_p}
\end{equation}
Taking advantage of the nilpotent nature of odd coordinates, we can collect
the terms in the formal series into a superfield
\begin{equation}\label{supercampotpm}
f(Y,\hbar)=\phi(y^m,\hbar)+\eta^\alpha \xi_\alpha(y^m,\hbar)+\cdots,
\end{equation}
whose components are elements of a standard Weyl algebra
\begin{equation}\label{campotpm}
\phi(y^m,\hbar)=\sum_{k,p}\hbar^k\; \phi_{k,m_1,\ldots,m_p} y^{m_1}\cdots
y^{m_p}.
\end{equation}
The component fields in \eqref{campotpm} will have tensor or spinor indices, and
Graßmann parity inherited from coefficients in the formal series
\eqref{supercampotpm}. An equivalent description is obtained by
taking a formal series in even coordinates
\begin{equation}
f(Y,\hbar)=\sum_{k,p}\hbar^k\; f_{k,m_1,\ldots,m_p} y^{m_1}\cdots y^{m_p},
\end{equation}
whose coefficients are superfields of tangent space odd coordinates.
\begin{equation}
f_{k,m_1,\ldots,m_p}=\phi_{k,m_1,\ldots,m_p}+\eta^\alpha
\xi_{\alpha,k,m_1,\ldots,m_p}+\cdots
\end{equation}

The vector bundle corresponding to the sheaf of Weyl superalgebras is the
\emph{Weyl superalgebra bundle} $\mathcal{W}$, whose sections are functions of
the symplectic base space as in 
\begin{equation}
f(z,Y,\hbar)=\sum_{k,p}\hbar^k\; f_{k,A_1,\ldots,A_p}(z) Y^{A_1}\cdots Y^{A_p},
\end{equation}
or equivalently, superfields like \eqref{supercampotpm} with components having
non constant coefficients $f_{k,m_1,\ldots,m_p}(z)$.

A Weyl superalgebra can be then defined by means of the following  Weyl
product,
\begin{equation}
f\circledcirc g=\left.\exp\left( -\frac{\iim \hbar}2\omega^{MN}
\frac\partial{\partial Y^M\vphantom{\partial\tilde{Y}^N}}
\frac\partial{\partial\tilde{Y}^N}\right)
f(Y,\hbar)g(\tilde{Y},\hbar)\right|_{\tilde{Y}=Y}.
\label{superWeyl}
\end{equation}
Since this product should preserve the degree, we propose $\deg(\eta)=1$.
Similar structures that appeared in the context of anti-Poisson brackets on
Graßmann algebras \cite{Bayen:1977hb} were developed further into a 
Fedosov-type procedure on the dual Graßmann algebra bundle of
a given vector bundle \cite{Bordemann:1996kd,Bordemann:1999ca}.
Note that the product \eqref{superWeyl} differs from such anti-Moyal
products as it acts on the tangent superspace $T(\mathcal{M})$, thus protecting 
super Poincaré symmetry on $\mathcal{M}$ from breaking. In
Darboux coordinates \cite{Rothstein:1990bj, Khudaverdian:2000zt} the Poisson
structure $\omega^{MN}$ becomes block diagonal allowing a separation of the sum
inside the exponential into a purely bosonic and a purely fermionic part. Each
term in turn, as a bilinear operator, is even and commutes with the others. For
instance
\begin{equation}
\left[\omega^{mn}\frac\partial{\partial y^m}\frac\partial{\partial
\tilde{y}^{n}},
\omega^{ab}\frac\partial{\partial \eta^a}\frac\partial{\partial
\tilde{\eta}^b}
\right]=0
\end{equation}
which is also true for any other combination of indices (note that a term with
mixed indices $\omega^{ma}$ is absent). This allow us to split the
exponential 
\begin{multline}
\exp\left( -\frac{\iim \hbar}2\omega^{MN}
\frac\partial{\partial Y^M\vphantom{\partial\tilde{Y}^N}}
\frac\partial{\partial\tilde{Y}^N}\right)=\\
\exp\left( -\frac{\iim \hbar}2\omega^{mn}
\frac\partial{\partial y^m}\frac\partial{\partial \tilde{y}^n}\right)
\exp\left( -\frac{\iim \hbar}2\omega^{ab}
\frac\partial{\partial \eta^a}\frac\partial{\partial \tilde{\eta}^b}\right),
\end{multline}
and expand the Weyl superproduct as the nested application of two products: The
usual Weyl rule for standard spacetime,
\begin{equation}
f\circ g=\left.\exp\left( -\frac{\iim \hbar}2\omega^{mn}
\frac\partial{\partial y^m}\frac\partial{\partial \tilde{y}^n}\right)
f(Y,\hbar)g(\tilde{Y},\hbar)\right|_{\tilde{Y}=Y}
\end{equation}
and a new one for the fermionic coordinates
\begin{equation}
f\bigcirc g=\left.\exp\left( -\frac{\iim \hbar}2\omega^{ab}
\frac\partial{\partial \eta^a}\frac\partial{\partial \tilde{\eta}^b}\right)
f(Y,\hbar)g(\tilde{Y},\hbar)\right|_{\tilde{Y}=Y}.
\end{equation}
Though both this products are formal series of the exponential
\begin{equation}
f\circ g=\sum_{k=0}^\infty \left( -\frac{\iim \hbar}2\right)^k
\frac1{k!}\omega^{m_1n_1}\ldots\omega^{m_kn_k}
\frac{\partial^kf}{\partial y^{m_1}\ldots \partial y^{m_k}}
\frac{\partial^kg}{\partial \tilde{y}^{n_1}\ldots \partial \tilde{y}^{n_k}},
\end{equation}
in the fermionic case the series is truncated due to nilpotency of odd
derivations. Nesting the products $\circ$ and $\bigcirc$ can be understood as
the replacement of the standard product in the formal series of $\circ$ with
the product rule of $\bigcirc$, that is
\begin{equation}
f\circledcirc g=\sum_{k=0}^\infty \left( -\frac{\iim \hbar}2\right)^k
\frac1{k!}\omega^{m_1n_1}\ldots\omega^{m_kn_k}
\frac{\partial^kf}{\partial y^{m_1}\ldots \partial y^{m_k}}
\bigcirc
\frac{\partial^kg}{\partial \tilde{y}^{n_1}\ldots \partial \tilde{y}^{n_k}}.
\end{equation}
Weyl products for the purely bosonic case (only $\omega^{mn}\neq0$) are
associative.  Including the bosonic and the full fermionic blocks into the form
$\omega^{MN}$ will lead to an associative product as a result of nesting the
$\bigcirc$ products already known to be associative. As the defined product is
covariant, a completely general proof of associativity can be worked out in
Darboux coordinates. This can be verified directly by expanding the product
$f\bigcirc(g\bigcirc h)$ for the fermionic part exclusively. One can see why
this work by looking at the particular case with only two fermionic coordinates
$Y^A=(y^m,\eta^1,\eta^2)$, which has a particular importance in the description
of the Weyl bundle over chiral superspace having precisely such
coordinates. The Weyl rule in this setup will looks like 
\begin{equation}
	f\bigcirc g =f g + \left( -\frac{\iim \hbar}2
\right)\omega^{\alpha\beta}(-1)^{p(f)}
	\partial_\alpha  f \partial_\beta g
	-\frac12\left( -\frac{\iim \hbar}2
\right)^2\omega^{\alpha\beta}\omega^{\gamma\delta}
	\partial_{\alpha\gamma}f \partial_{\beta\delta} g,
\label{ChiralWeylRule}
\end{equation}
where we use the standard spinor index notation for the odd variables
$\alpha=1,2$ and the usual shorthand $\partial_\alpha=
\frac{\partial}{\partial\eta^\alpha}$ to denote derivation respect to
$Y^\alpha=\eta^\alpha$ (with its straightforward extension
$\partial_{\alpha\beta}= \partial_\alpha\partial_\beta)$. Note that
the series is truncated to second order, which facilitates direct verification
of the associativity
\begin{multline}
	(f \bigcirc g)\bigcirc h=f g h\\
        +\left( -\frac{\iim \hbar}2\right)\omega^{\alpha\beta}
	\bigl[(-1)^{p(fg)}\partial_\alpha f g \partial_\beta  h
            +(-1)^{p(g)} f \partial_\alpha g \partial_\beta h
        +(-1)^{p(f)}\partial_\alpha f \partial_\beta g h
	\bigr]\\
	-\frac12\left( -\frac{\iim \hbar}2\right)^2
        \omega^{\alpha\beta}\omega^{\gamma\delta}
	\bigl[\partial_{\alpha\gamma}f g \partial_{\beta\delta}h
	+f \partial_{\alpha\gamma}g \partial_{\beta\delta}h
	+\partial_{\alpha\gamma}f \partial_{\beta\delta}g h\\
	-2(-1)^{p(fg)}\partial_\alpha f \partial_{\beta\gamma}g \partial_\delta
h
	-2(-1)^{p(g)} \partial_{\gamma\alpha}f \partial_\beta g \partial_\delta
h
	+(-1)^{p(f)}\partial_\alpha f \partial_\gamma g
       \partial_{\beta\delta}h
	\bigr]\\
        =f\bigcirc (g\bigcirc h).
\end{multline}
A $\bigcirc$ product involving $4N$ odd variables, as required by more general
$N$--supersymmetric theories, is obtained by repeatedly nesting products as
\eqref{ChiralWeylRule} above, resulting in an associative product with a
truncated expansion of order $4N$. If we include the standard $\circ$ product in
Darboux coordinates one also concludes that this product is associative in
general, from the covariant character of $\cc$. 

The Weyl superalgebra is easily extended to include $q$-superforms on
$\mathcal{M}$, that is, sections on ${\mathcal{W}}\otimes
\Omega^*$
expressed as
\begin{equation}
f(z,Y,\hbar)=\sum_{k,p}\hbar^k\; f_{k,A_1,\ldots,A_p,B_1\ldots,B_q}(z)
Y^{A_1}\cdots Y^{A_p}
dz^{B_1}\wedge\ldots\wedge dz^{B_q}.
\end{equation}
These differential forms constitute an algebra with multiplication defined by
the combined exterior superproduct of forms given in \S\ref{ssm} together with
the Weyl product acting on tangent superspace variables. 

The commutator of 2 superforms $f\in {\mathcal{W}}\otimes \Omega^{q_1}$ and
$g\in {\mathcal{W}}\otimes \Omega^{q_2}$  is   
\begin{equation}\label{corchete}
\left[f,g\right]=f\circledcirc g-(-1)^{q_1 q_2}(-1)^{p(f)p(g)}g\circledcirc
f .
\end{equation}
A central superform  $f$ is such that for any $g\in {\mathcal{W}}\otimes
\Omega^*$, $\left[f,g\right]=0$. In \eqref{corchete} the wedge exterior product
$\wedge$ acting on forms is understood.

\section{Superspace Deformations}
\subsection{Abelian subalgebra and deformations}
In the purely bosonic case the Weyl product is linked to deformations on the
base space through a correspondence between $C^{\infty}(M)$ and the Abelian
subalgebra that, in Darboux coordinates, results in the well known Moyal
product, as was proved in \cite{Fedosov:1996fu}. In this section we will show
that this is also the case for the  Weyl superproduct, which
for the chiral case give us precisely the
nonanticonmutative product extensively studied in this decade, see for example
\cite{Seiberg:2003yz, Ferrara:2003xy, Klemm:2001yu}. 
      
We start with the concept of Abelian connection. 
A connection $\Dc_A$ on the Bundle $\mathcal{W}$ is called
{\it{Abelian}} if for any section $f\in \mathcal{W}\otimes\Omega^*$
\begin{equation}
\mathcal {D}_A^2 f=0
\end{equation}

Inspired by Fedosov construction, we will lift the supersymplectic connection
defined in \S \ref{ssm} to the Weyl superalgebra bundle
\begin{equation}
\Dc=dz^A\wedge \nabla_A f.
\end{equation}
$\Dc$ is a covariant derivative acting on $C^\infty(\mathcal{W}\otimes
\Omega^*)$ and obeying the following properties
\begin{eqnarray*}
&\Dc(f\cc g)=\Dc f \cc g +
(-1)^{q_1+p(f)}f\cc\Dc g\\
&\Dc\left[f\;,\;g\right]=\left[\Dc f, g\right] +
(-1)^{q_1+p(f)}\left[f, \Dc g\right]
\end{eqnarray*}

Another important operators involved in the construction of the Abelian
connection are $\delta: \mathcal{W}_p\otimes \Omega^q \rightarrow
\mathcal{W}_{p-1}\otimes \Omega^{q+1}$ that raises the rank of forms and acts as
a sort of exterior superderivative, and $\delta^*: \mathcal{W}_p\otimes \Omega^q
\rightarrow \mathcal{W}_{p+1}\otimes \Omega^{q-1}$, that lowers the rank of
forms and acts as a contraction operator. In coordinate basis they are realized
as
\begin{align}
\delta f= dz^A\wedge \frac\partial{\partial Y^A} f && \delta^* f= \iim
Y^A\wedge \frac\partial{\partial z^A} f,
\end{align}
Using \eqref{zz}, \eqref{dzdz}  and 
\begin{eqnarray}
dz^A\wedge Y^B =(-1)^{p(Y)p(dz)},
\end{eqnarray}
and observing their action on each term in the formal series 
\begin{equation}
f_{k,A_1,\ldots,A_p, B_1,\ldots, B_q}Y^{A_1}\cdots Y^{A_p}dz^{B_1}\wedge \cdots
\wedge dz^{B_q},
\end{equation}
one can check their properties
\begin{align}
&\delta^2=0, \quad\quad (\delta^*)^2=0,\\
&\delta(f\circledcirc g)=
\delta f \circledcirc  g
+(-1)^{p(f)+q} f\circledcirc\delta g,\\
&f =\delta \delta^{-1} f + \delta^{-1} \delta f +f_{00}.
\end{align}
Here $f_{pq}$ is the coefficient with $p$ times $Y$ and $q$ times $dz$, so
that  $(p+q)\delta^{-1}f_{pq}= \delta^* f_{pq}$. 

Given the supersymplectic connection $\Dc$, there always exists an Abelian
connection
\begin{equation}
\Dc_A\;\cdot= \Dc \cdot 
+\frac\iim\hbar
\left[(\omega_{AB}Y^AY^B+r),\;\cdot\;\right]
\end{equation}
where $r\in C^\infty(\mathcal{W}_3\otimes\Omega^1)$ satisfies $\delta^{-1}r=0$.

The proof follows the corresponding one in Fedosov \cite[\S5.2]{Fedosov:1996fu},
where mainly the properties of $\delta$, $\delta^{-1}$ and degree counting are
used. The arguments are exactly the same. The curvature of the Abelian
connection is
\begin{equation}
F=-\frac12\omega_{AB}\,dz^A\wedge dz^B+R-\delta r
+\Dc r+\frac\iim\hbar r^2
\end{equation}
We consider then $r$ satisfying
\begin{equation}
\delta r=R+\Dc r+\frac\iim\hbar r^2,
\label{deltar}
\end{equation}
which ensures that $\Dc_A$ is Abelian. Eq \eqref{deltar} has a unique solution
belonging to $C^\infty(\mathcal{W}_3\otimes\Omega^1)$ and satisfying
$\delta^{-1}r=0$. The main point in Fedosov's proof, which can be extended to
the supersymplectic case, is to show that
\begin{equation}
r=\delta^{-1}\left(R+\Dc r+\frac\iim\hbar r^2\right)
\end{equation}
has a unique solution
belonging to $C^\infty(\mathcal{W}_3\otimes\Omega^1)$. This follows from the
iteration procedure defined by
\begin{equation}
r_{n+1}=\delta^{-1}R+\delta^{-1}\left(\Dc r_n+\frac\iim\hbar r_n^2\right),
\qquad n=0,1,2,\ldots
\end{equation}
From degree counting one obtains $\deg(r_n-r_{n+1})\ge n+3$ implying that the
terms for any fixed degree of the sequence are the same for large enough $n$.
There exists then a unique solution in
$C^\infty(\mathcal{W}_3\otimes\Omega^1)$. It is then straightforward to show
that this solution satisfies \eqref{deltar} as $\delta^{-1}r=0$ holds by
construction. Uniqueness arises from acting on \eqref{deltar} with
$\delta^{-1}$ and using the above property of $r$.

With this at hand we propose an Abelian connection of the simplest form 
\begin{equation}
\Dc_A f= \Dc f - \delta f
\end{equation}
Once we have the Abelian connection, we use it to find the elements of the
Abelian sub(super)algebra $\mathcal{W}_A= \left\{f\in \mathcal{W}: \Dc_A
f=0\right \}$.  There is a one-to-one correspondence between the set of
solutions to the equation $\Dc_A f=0$ and functions on the base space
$\mathcal{M}$, that relates the Weyl product to a deformation on the base space.
A proof of this correspondence is a straightforward
generalization of the original theorem by Fedosov (5.2.4 in
\cite{Fedosov:1996fu}).
 
Let us consider a section $a\in C^\infty(\mathcal{W})$, $a=a(x,\theta,y,\eta)$,
and focus in the Abelian subalgebra $\Dc_A a=0$. As in Fedosov
\cite{Fedosov:1996fu} we take
\begin{equation}
\delta a=(\Dc_A+\delta)a,
\end{equation}
being $a$ a 0--form for which $\delta^{-1}a=0$ holds, the Hodge-De Rham
decomposition results in
\begin{equation}
a=a|_{y=\eta=0}+\delta^{-1}(\Dc_A+\delta)a.
\label{HdRdecomp}
\end{equation}
Given $a|_{y=\eta=0}=a(x,\theta,0,0)$, as $\delta^{-1}$ raises degree, an
iteration procedure yields existence and uniqueness of the solution. The
bijection then follows from \eqref{HdRdecomp}.

Considering bosonic coordinates only, it is known
that the deformation induced in base space by a Weyl product in
Darboux coordinates is precisely the Moyal product \cite{Fedosov:1996fu,
Martin:2001zv}. More succinctly, let $\sigma:\mathcal{W}_A \longrightarrow
C^\infty(\mathcal{M})[[\hbar]]$ be the projection to the center that maps
$\sigma:f\mapsto f_{00}$ and
$\sigma:g\mapsto g_{00}$, then
\begin{equation}
\sigma(f\circ g)=f_{00}\star g_{00},
\end{equation}
where $\star$ stands for the Moyal product. In the next section we will obtain
the nonanticommutative $\star$ product, proposed in \cite{Seiberg:1999vs,
Ferrara:2003xy, Klemm:2001yu} as a projection of the $\cc$ superproduct to the
base space using Darboux chiral coordinates in superspace.

\section{Nonanticommutative $Q$--deformations from Weyl superproduct}
\label{NonAntiSection}
 In this section we will restrict ourselves to functions
$f\in C^\infty(\mathcal{W}\otimes \Omega^0)$ whose coefficients are chiral
superfields, i.e. $\bar{D}^{\alphad}f=0$, where $\bar{D}^{\alphad}$ stands for
the supersymmetric derivative in the base space. In
order to obtain nonanticommutativity in the base space, we will drop 
$\omega^{mn}$ and $\omega^{\alphad\betad}$ terms. The Weyl rule is directly
generalized to this setup in Darboux coordinates 
\begin{equation}
	f\cc g=f\bigcirc g =fg + \left( -\frac{\iim \hbar}2
\right)\omega^{\alpha\beta}(-1)^{p (f)}
	\partial_\alpha f \partial_\beta g
	-\frac12\left( -\frac{\iim \hbar}2
\right)^2\omega^{\alpha\beta}\omega^{\gamma\delta}
	\partial_{\alpha\gamma}f \partial_{\beta\delta}g,
\end{equation}
Now we consider the following Abelian connection
\begin{equation}
\Dc_A f=\Dc  f-\delta f=\Dc f +\frac\iim{\hbar}
\left[\omega_{\alpha\beta}\eta^\alpha d\theta^\beta, f \right],
\end{equation}
and build the Abelian sub algebra of Graßmann-even superfields from the
condition $\Dc_A f=0$. The symplectic connection can be build using the
super Poincaré generator
\begin{equation}
\Dc=d\theta^\alpha\wedge Q_\alpha
\end{equation}
in chiral coordinates, we have
\begin{equation}
\Dc_Af=d\theta^\alpha\wedge\frac\partial{\partial\theta^\alpha}f
+\omega^{\alpha\gamma}\omega_{\alpha\beta}d\theta^\beta
\frac\partial{\partial\eta^\gamma}f
\end{equation}
Focusing on Graßmann-even superfields on the generalization of the Weyl bundle
\begin{equation}
	f=f_0+\hbar
f_{1\alpha}\eta^\alpha+\hbar^2f_{2\alpha\beta}\eta^\alpha\eta^\beta,
\end{equation}
the Abelian condition reduces to the following set of equations
\begin{equation}
   \begin{aligned}
      &\frac{\partial f_0}{\partial\theta^\gamma}
      +h\omega^{\alpha\beta}\omega_{\alpha\gamma}f_{1\beta}=0,\\
      &\frac{\partial f_{1\alpha}}{\partial\theta^\gamma}	
      +\hbar\omega^{\delta\beta}\omega_{\delta\gamma}
      2f_2\varepsilon_{\beta\alpha}=0,\\
      &\frac{\partial f_2}{\partial\theta^\gamma}=0.
   \end{aligned}
\label{AbelianConditionComponents1}
\end{equation}

As the symplectic form is everywhere not degenerate,
$\omega^{\alpha\beta}\omega_{\alpha\gamma}=\delta^\beta_\gamma$, we easily find
the general form of an element of the Abelian subalgebra to be
\begin{equation}
	f=f_0-Q_\alpha f_0\eta^\alpha+\frac14Q^\alpha Q_\alpha f_0
(\eta^\beta\eta_\beta)
\end{equation}

The product of two elements of this subalgebra is projected back to the
nonanticommutative Moyal product \cite{Seiberg:1999vs, Ferrara:2003xy,
Klemm:2001yu}
\begin{equation}
	f\bigcirc g=f_0g_0 
   -\frac{\iim \hbar}2\omega^{\alpha\beta}
   Q_\alpha f_0 Q_\beta g_0
	-\frac12\left( -\frac{\iim \hbar}2 \right)^2 \omega^2
   Q^\alpha Q_\alpha f_0 Q^\beta Q_\beta g_0
	+\Oh(\eta),
\label{NAC-Product}
\end{equation}
\begin{equation}
\sigma(f\bigcirc g)=f_0\star g_0.
\end{equation}

A first attempt to introduce curvature in the previous setup is to include
nonzero connection coefficients $\Gamma^\gamma_{\alpha\beta}$
\begin{equation}
(\nabla_\eta \omega)(\chi,\psi)=\eta^\alpha\chi^\beta\psi^\gamma
(\partial_\alpha\omega_{\beta\gamma}
+\Gamma_{\gamma\alpha\beta}+\Gamma_{\beta\alpha\gamma}).
\end{equation}
for Graßmann vector fields $\eta,\chi$ and $\psi$. i.e.
\begin{equation}
\eta=\eta^\alpha\frac{\partial}{\partial\theta^\alpha}
\end{equation}
The antisymmetric connection coefficients are defined up to an
arbitrary \emph{completely antisymmetric} tensor $\Theta_{\alpha\beta\gamma}$.

On  sections $a$ of the superbundle, the connection is lifted to
\begin{equation}
\Dc a=d\theta^\alpha\wedge \nabla_\alpha a=da+\frac{\iim}{h}[\Theta,a],
\end{equation}
with
\begin{equation}
\Theta=\frac12\Theta_{\alpha\beta\gamma}\eta^\alpha\eta^\beta d\theta^\gamma.
\end{equation}

An Abelian connection can then be constructed via
\begin{equation}
\Dc_Aa=da +\frac\iim{h}
\left[\omega_{\alpha\beta}\eta^\alpha d\theta^\beta
+\frac12\Theta_{\alpha\beta\gamma}\eta^\alpha\eta^\beta d\theta^\gamma,
a \right]_\circ.
\end{equation}
The associated Weyl subalgebra is obtained following the same procedure as for
\eqref{AbelianConditionComponents1}, leading to an analogous correspondence
between undeformed sections and Abelian ones, now including a more general
covariant derivative. The projection of the Weyl product results in an
expression which is formally the same as before \eqref{NAC-Product} due to
nilpotency of the Grassmann basis, which truncates the series to second order in
the deformation, preventing the appearance of terms involving the curvature.
This product does not correspond to the standard Moyal product, as it includes
connection coefficients. Explicit expressions for more general
non(anti)commutative products following the approach introduced in this work
will be presented in a forthcoming paper, this includes the description of
$D$--deformations.

\begin{acknowledgments}
This work was partially supported by the USB-DID S1 grants (A. De Castro and L.
Quevedo). A. Restuccia would like to thank the Max Planck Institut für
Gravitationsphysik for its hospitality.
\end{acknowledgments}

\appendix

\section{Conventions}
Spinor algebra follows the convention in Wess-Bagger
\begin{equation}
\begin{aligned}
\eta^\alpha\eta^\beta&=-\frac12\varepsilon^{\alpha\beta}(\eta\eta),
\qquad
\eta_\alpha\eta_\beta&=+\frac12\varepsilon_{\alpha\beta}(\eta\eta),\\
\bar{\eta}_{\dot{\alpha}}\bar{\eta}_{\dot{\beta}}
&=-\frac12\varepsilon_{\dot{\alpha}\dot{\beta}}(\bar{\eta}\bar{\eta}),\qquad
\bar{\eta}^{\dot{\alpha}}\bar{\eta}^{\dot{\beta}}
&=+\frac12\varepsilon^{\dot{\alpha}\dot{\beta}}(\bar{\eta}\bar{\eta}).
\end{aligned}
\end{equation}

But with a more comfortable definition for the derivative
\begin{equation}
\begin{aligned}
   \partial_\alpha&=\frac{\partial}{\partial\eta^\alpha},\qquad
   &\bar{\partial}^{\dot{\alpha}}&
   =\frac{\partial}{\partial\bar{\eta}_{\dot{\alpha}}},\\
   \partial^\alpha&=-\frac{\partial}{\partial\eta_\alpha}
   =\varepsilon^{\alpha\beta}\partial_\beta,\qquad
   &\bar{\partial}_{\dot{\alpha}}&=
   -\frac{\partial}{\partial\bar{\eta}^{\dot{\alpha}}}
   =\varepsilon_{\dot{\alpha}\dot{\beta}}\bar{\partial}^{\dot{\beta}}.
\end{aligned}
\end{equation}



\end{document}